\documentclass[a4paper,aps,twocolumn,nofootinbib]{revtex4}
\RequirePackage[colorlinks,hyperindex]{hyperref}
\RequirePackage[english]{babel}
\RequirePackage[latin1]{inputenc}
\RequirePackage[T1]{fontenc}
\RequirePackage{mathrsfs}
\RequirePackage{amsmath}
\RequirePackage{amssymb}
\RequirePackage{amsbsy}
\RequirePackage{color}
\RequirePackage{bm}
\hypersetup{colorlinks=true,breaklinks=true,urlcolor=blue,linkcolor=red}
\begin{document}
%%%%%%%%%%%%%%%%%%%%%%%%%%%%%%%%%%%%%%%%%%%%%%%%%%%%%%%%%%%%%%%%%%%%%%%%%%%%%%%%%%%%%%%%%%%%%%%%%%%
\title{\bf{A Square-Integrable Spinor Solution to Non-Interacting Dirac Equations}}
\author{Luca Fabbri, Roberto Cianci, Stefano Vignolo}
\affiliation{DIME, Sez. Metodi e Modelli Matematici, Universit\`{a} di Genova, 
Via all'Opera Pia 15, 16145 Genova, ITALY}
\date{\today}
%%%%%%%%%%%%%%%%%%%%%%%%%%%%%%%%%%%%%%%%%%%%%%%%%%%%%%%%%%%%%%%%%%%%%%%%%%%%%%%%%%%%%%%%%%%%%%%%%%%
\begin{abstract}
We consider the Dirac equation written in polar form, without any external potential but equipped with a non-zero tensorial connection, and we find a new type of solution that is localized around the origin with a decreasing exponential behaviour in the radial coordinate.
\end{abstract}
%%%%%%%%%%%%%%%%%%%%%%%%%%%%%%%%%%%%%%%%%%%%%%%%%%%%%%%%%%%%%%%%%%%%%%%%%%%%%%%%%%%%%%%%%%%%%%%%%%%
\maketitle
%%%%%%%%%%%%%%%%%%%%%%%%%%%%%%%%%%%%%%%%%%%%%%%%%%%%%%%%%%%%%%%%%%%%%%%%%%%%%%%%%%%%%%%%%%%%%%%%%%%
\section{Introduction}
One of the most important problems in mathematical physics is to study the properties of the Dirac spinor field equations by assigning a specific potential and finding its exact solutions, according to various methods \cite{Bargov1977, Trevisan2014, Brandt2016, Oliveira2019, CS, CP, GM1, GM2, shishkin1989dirac, shishkin1989dirac2, villalba1994angular, villalba2002separation, kruger1991new}.

Whereas the presence of specific forms of external potentials provides properties of the matter distributions in those potentials, studying the non-interacting case would allow to deduce properties about the matter distributions themselves in the cleanest way.\!\! Nonetheless, it is generally believed that in the free case we can only get solutions such as plane waves, or similarly non-integrable solutions.

Such a belief is based on the argument that the lack of external potentials is equivalent to the total lack of force that can keep the matter distribution localized. Nonetheless, non-interacting does not necessarily mean to be fully free. In fact, the above argument fails to contemplate the situation for which some matter distributions might have some form of internal tensions that can keep them localized even when no source of any type is to be around.

Mathematically, this can be seen in the fact that Dirac spinor field equations are defined in terms of frames, that is objects that form the soldering between diffeomorphic and Lorentz structures. As it is known, spinors are more sensitive to the structure of space-time than real tensors \cite{Steane:2013wra} while frames can contain more information than the metric, it follows that the interplay between spinors and frames can provide additional information than we would normally think. By translating everything in polar form, the form in which spinor components are written as modules times phases while respecting covariance, frames are translated in objects called tensorial connections, and in terms of which the above consideration is clearer. In fact, the tensorial connections can be seen as the potential of the Riemann curvature while being a real tensor, and as a zero Riemann curvature can still be obtained in terms of non-zero tensorial connection, such tensorial connection describes a space-time structure where no external force is present but which nevertheless is not free. In this case, solutions can be found displaying the character of radial localization, as we are going to discuss in this work.
%%%%%%%%%%%%%%%%%%%%%%%%%%%%%%%%%%%%%%%%%%%%%%%%%%%%%%%%%%%%%%%%%%%%%%%%%%%%%%%%%%%%%%%%%%%%%%%%%%%
%%%%%%%%%%%%%%%%%%%%%%%%%%%%%%%%%%%%%%%%%%%%%%%%%%%%%%%%%%%%%%%%%%%%%%%%%%%%%%%%%%%%%%%%%%%%%%%%%%%
\section{Spinor Fields in Polar Form}
We will use units $c\!=\!\hbar\!=\!1$ throughout the paper.

All along, Clifford matrices are given by $\boldsymbol{\gamma}^{a}$ such that $\left\{\boldsymbol{\gamma}_{a}\!,\!\boldsymbol{\gamma}_{b}\right\}\!=\!2\eta_{ab}\mathbb{I}$ with $\eta_{ab}$ being the Minkowski matrix and $\left[\boldsymbol{\gamma}_{a}\!,\!\boldsymbol{\gamma}_{b}\right]\!=\!4\boldsymbol{\sigma}_{ab}$ will define the infinitesimal generators of a complex Lorentz algebra (in this paper, we specify onto the spin-$1/2$ representation) while $2i\boldsymbol{\sigma}_{ab}\!=\!\varepsilon_{abcd}\boldsymbol{\pi}\boldsymbol{\sigma}^{cd}$ defines the $\boldsymbol{\pi}$ matrix (this matrix is usually denoted as a gamma matrix with an index five, but since in space-time this index has no meaning, and sometimes it may also be misleading, we use a notation with no index).\! Tetrads $e^{a}_{\nu}$ and metric $g_{\mu\nu}$ will describe the metric of the space-time.

With gamma matrices, and the pair of adjoint spinors $\psi$ and $\overline{\psi}$, we construct the bi-linear spinor quantities
\begin{eqnarray}
&S^{a}\!=\!\overline{\psi}\boldsymbol{\gamma}^{a}\boldsymbol{\pi}\psi\\
&U^{a}\!=\!\overline{\psi}\boldsymbol{\gamma}^{a}\psi\\
&\Theta\!=\!i\overline{\psi}\boldsymbol{\pi}\psi\\
&\Phi\!=\!\overline{\psi}\psi
\end{eqnarray}
which are all real tensors \cite{HoffdaSilva:2019xvd}. If $\Theta$ and $\Phi$ are not at the same time equal to zero identically, we can always write the spinor field, in chiral representation, in the form
\begin{eqnarray}
&\!\psi\!=\!\phi e^{-\frac{i}{2}\beta\boldsymbol{\pi}}
\boldsymbol{S}\left(\!\begin{tabular}{c}
$1$\\
$0$\\
$1$\\
$0$
\end{tabular}\!\right)
\label{spinorch}
\end{eqnarray}
for some complex Lorentz transformation $\boldsymbol{S}$ with $\phi$ and $\beta$ called module and Yvon-Takabayashi angle, and where we can appreciate the polar form of each component and the manifest general Lorentz covariance \cite{L,Cavalcanti:2014wia,j-l,Fabbri:2016msm}. By considering the polar form of the spinor field, we have that
\begin{eqnarray}
&S^{a}\!=\!2\phi^{2}s^{a}\\
&U^{a}\!=\!2\phi^{2}u^{a}\\
&\Theta\!=\!2\phi^{2}\sin{\beta}\\
&\Phi\!=\!2\phi^{2}\cos{\beta}
\end{eqnarray}
restricted by $u_{a}u^{a}\!=\!-s_{a}s^{a}\!=\!1$ and $u_{a}s^{a}\!=\!0$ and showing that module and Yvon-Takabayashi angle are one scalar and one pseudo-scalar and the only degrees of freedom of the spinor field. In fact, the advantage of writing spinors in polar form is that the $8$ real components are rearranged into that special configuration in which the $2$ real scalar degrees of freedom ($\phi$ and $\beta$) remain isolated from the $6$ real components (the spatial parts of $u_{a}$ and $s_{a}$) that can always be transferred into the frame. This also gives the interpretation of these mathematical objects, since the $3$ spatial parts of $u_{a}$ can always be vanished by means of $3$ boosts and therefore $u_{a}$ is the velocity vector. Similarly, the $3$ spatial parts of $s_{a}$ can always be vanished by means of $3$ rotations and so $s_{a}$ is the spin axial-vector (another way to prove this is to see that $s_{a}$ is proportional to the Hodge dual of the completely antisymmetric spin of the spinor field \cite{j-l,Fabbri:2016msm}). Borrowing for the case of the plane waves, or in non-relativistic approximations, one can see that the module $\phi$ represents the overall distributions of the matter field. The Yvon-Takabayashi angle $\beta$ is simply the phase difference of the two chiral parts, and as such it represents some form of internal dynamics (alternatively one can see that $\beta$ is a degree of freedom about internal dynamics because if non-zero it forbids a non-relativistic limit even in the frame at rest of the spinor field \cite{Fabbri:2016msm}).

Writing the covariant derivative of the spinor in polar form, we can prove that it is always possible to write
\begin{eqnarray}
&\!\!\!\!\!\!\!\!\boldsymbol{\nabla}_{\mu}\psi\!=\!(-\frac{i}{2}\nabla_{\mu}\beta\boldsymbol{\pi}
\!+\!\nabla_{\mu}\ln{\phi}\mathbb{I}
\!-\!iP_{\mu}\mathbb{I}\!-\!\frac{1}{2}R_{ij\mu}\boldsymbol{\sigma}^{ij})\psi
\label{decspinder}
\end{eqnarray}
in terms of a vector $P_{\mu}$ and a tensor $R_{ij\mu}$ called \emph{tensorial connection} because it encodes all information about both gravity and effects related to non-inertial frames \cite{Fabbri:2018crr}.

It is also possible to have the Riemann curvature tensor written in terms of the tensorial connection according to
\begin{eqnarray}
&\!\!\!\!\!\!\!\!\!\!\!\!R^{i}_{\phantom{i}j\mu\nu}\!=\!-(\nabla_{\mu}R^{i}_{\phantom{i}j\nu}
\!-\!\!\nabla_{\nu}R^{i}_{\phantom{i}j\mu}
\!\!+\!R^{i}_{\phantom{i}k\mu}R^{k}_{\phantom{k}j\nu}
\!-\!R^{i}_{\phantom{i}k\nu}R^{k}_{\phantom{k}j\mu})\label{Riemann}
\end{eqnarray}
with the tensorial connection encoding information about gravity and frames but with the curvature containing information about gravity alone. If we could find non-zero solutions of equations (\ref{Riemann}) after setting the Riemann curvature to zero, they would represent tensorial connections encoding information related to frames solely \cite{Fabbri:2018crr}. An example in spherical coordinates is given according to the
\begin{eqnarray}
&R_{r\varphi\varphi}\!=\!-r(\sin{\theta})^{2}\label{1}\\
&R_{\theta\varphi\varphi}\!=\!-r^{2}\cos{\theta}\sin{\theta}\label{2}
\end{eqnarray}
with
\begin{eqnarray}
&R_{rtt}\!=\!-2\varepsilon\sinh{\alpha}\sin{\rho}\label{3}\\
&R_{\varphi rt}\!=\!2\varepsilon r\sin{\theta}\cosh{\alpha}\sin{\rho}\label{4}\\
&R_{\theta tt}\!=\!2\varepsilon r\sinh{\alpha}\cos{\rho}\label{5}\\
&R_{\varphi\theta t}\!=\!-2\varepsilon r^{2}\sin{\theta}\cosh{\alpha}\cos{\rho}\label{6}
\end{eqnarray}
and
\begin{eqnarray}
&r\sin{\theta}\partial_{\theta}\alpha\!=\!R_{t\varphi\theta}\label{7}\\
&r\sin{\theta}\partial_{r}\alpha\!=\!R_{t\varphi r}\label{8}\\
&-r(1\!+\!\partial_{\theta}\rho)\!=\!R_{r\theta\theta}\label{9}\\
&r\partial_{r}\rho\!=\!R_{\theta rr}\label{10}
\end{eqnarray}
in terms of two functions $\alpha\!=\!\alpha(r,\theta)$ and $\rho\!=\!\rho(r,\theta)$ still not specified while the constant $\varepsilon$ has the meaning of an energy describing some tension of the vacuum \cite{Fabbri:2019kfr}.

For the dynamics, we take the spinor field subject to
\begin{eqnarray}
&i\boldsymbol{\gamma}^{\mu}\boldsymbol{\nabla}_{\mu}\psi\!-\!m\psi\!=\!0
\label{Dirac}
\end{eqnarray}
called Dirac equation \cite{p-s}. In polar form it is given by
\begin{eqnarray}
&\!\!\!\!B_{\mu}\!-\!2P^{\iota}u_{[\iota}s_{\mu]}\!+\!\nabla_{\mu}\beta
\!+\!2s_{\mu}m\cos{\beta}\!=\!0\label{dep1}\\
&\!\!\!\!R_{\mu}\!-\!2P^{\rho}u^{\nu}s^{\alpha}\varepsilon_{\mu\rho\nu\alpha}
\!+\!2s_{\mu}m\sin{\beta}\!+\!\nabla_{\mu}\ln{\phi^{2}}\!=\!0\label{dep2}
\end{eqnarray}
with $R_{\mu a}^{\phantom{\mu a}a}\!=\!R_{\mu}$ and $\frac{1}{2}\varepsilon_{\mu\alpha\nu\iota}R^{\alpha\nu\iota}\!=\!B_{\mu}$ \cite{Fabbri:2016laz,Fabbri:2019kyd}. We notice that (\ref{Dirac}) are $8$ real equations and thus as many as the $2$ vector equations (\ref{dep1}, \ref{dep2}), which specify all the space-time derivatives for both Yvon-Takabayashi angle and module of spinors \cite{Fabbri:2016laz}. Field equations (\ref{dep1}, \ref{dep2}) become
\begin{eqnarray}
\nonumber
&\!\!\!\!\!\!\!\!r\partial_{r}\beta\!+\!\partial_{\theta}\alpha-\\
&\!\!\!\!\!\!\!\!\!\!\!\!-[2(\varepsilon+E)r\cosh{\alpha}\!-\!2L\frac{\sinh{\alpha}}{\sin{\theta}}
\!-\!2mr\cos{\beta}]\cos{\rho}
\!=\!0\label{a}\\
\nonumber
&\!\!\!\!\!\!\!\!\partial_{\theta}\beta-r\partial_{r}\alpha-\\
&\!\!\!\!\!\!\!\!-[2(\varepsilon+E)r\cosh{\alpha}\!-\!2L\frac{\sinh{\alpha}}{\sin{\theta}}
\!-\!2mr\cos{\beta}]\sin{\rho}
\!=\!0\label{b}\\
\nonumber
&\!\!\!\!r\partial_{r}\ln{(\phi^{2}r^{2}\sin{\theta})}\!+\!2mr\sin{\beta}\cos{\rho}
\!+\!\partial_{\theta}\rho-\\
&\!\!\!\!-[2(\varepsilon+E)r\sinh{\alpha}\!-\!2L\frac{\cosh{\alpha}}{\sin{\theta}}]\sin{\rho}
\!=\!0\label{c}\\
\nonumber
&\!\!\!\!\partial_{\theta}\ln{(\phi^{2}r^{2}\sin{\theta})}\!+\!2mr\sin{\beta}\sin{\rho}
\!-\!r\partial_{r}\rho+\\
&\!\!\!\!+[2(\varepsilon+E)r\sinh{\alpha}\!-\!2L\frac{\cosh{\alpha}}{\sin{\theta}}]\cos{\rho}
\!=\!0\label{d}
\end{eqnarray}
if in parallel to (\ref{1}-\ref{10}) we also take
\begin{eqnarray}
&\!\!\!\!e_{0}^{t}\!=\!\cosh{\alpha}\ \ \ \ e_{2}^{t}\!=\!-\sinh{\alpha}\\
&\!\!\!\!e_{1}^{r}\!=\!\sin{\rho}\ \ \ \ e_{3}^{r}\!=\!-\cos{\rho}\\
&\!\!\!\!e_{1}^{\theta}\!=\!-\frac{1}{r}\cos{\rho}\ \ \ \ 
e_{3}^{\theta}\!=\!-\frac{1}{r}\sin{\rho}\\
&\!\!\!\!e_{0}^{\varphi}\!=\!-\frac{1}{r\sin{\theta}}\sinh{\alpha}\ \ \ \ 
e_{2}^{\varphi}\!=\!\frac{1}{r\sin{\theta}}\cosh{\alpha}
\end{eqnarray}
and
\begin{eqnarray}
&s_{r}\!=\!\cos{\rho}\ \ \ \ \ \ s_{\theta}\!=\!r\sin{\rho}\\
&u_{t}\!=\!\cosh{\alpha}\ \ \ \ \ \ u_{\varphi}\!=\!r\sin{\theta}\sinh{\alpha}
\end{eqnarray}
as well as
\begin{eqnarray}
&P_{t}\!=\!E\ \ \ \ \ \ \ \ P_{\varphi}\!=\!L
\end{eqnarray}
in which $E$ and $L$ are constants. In quantum field theory they are associated to the energy and angular momentum of the particle, and so they are $E\!=\!m$ and $L\!=\!\pm1/2$ for particles of rest mass $m$ and spin $1/2$ as considered here.
%%%%%%%%%%%%%%%%%%%%%%%%%%%%%%%%%%%%%%%%%%%%%%%%%%%%%%%%%%%%%%%%%%%%%%%%%%%%%%%%%%%%%%%%%%%%%%%%%%%
%%%%%%%%%%%%%%%%%%%%%%%%%%%%%%%%%%%%%%%%%%%%%%%%%%%%%%%%%%%%%%%%%%%%%%%%%%%%%%%%%%%%%%%%%%%%%%%%%%%
\section{New Trial Solutions}
When dealing with a field equation, one thing to do is to look for exact solutions. Because this task is generally quite difficult, one normally picks exact solutions of some specific form, such as for instance plane waves, so to ease the search. Because we have recalled how to re-write the Dirac equations in polar form, our next task would be to show what is the advantage of having the polar form in choosing a special type of exact solution. We will see in fact that, in polar form, it is much easier to guess a trial solution, more general than the ones we know so far.

In this last form (\ref{a}-\ref{d}), the Dirac field equations seem to be quite simple but also general enough to account for exact solutions of specific cases given by the integrable potentials known in physics, that is the Coulomb potential and the elastic potential. Nonetheless, these solutions can do much more, since they share a common structure, as it can be made manifest by comparing side to side the solution for the hydrogen atom \cite{Fabbri:2018crr} against the solution of the harmonic oscillator \cite{Fabbri:2019kyd}. Their common form has
\begin{eqnarray}
\sin{\rho}\!=\!\frac{X\sin{\theta}}{\sqrt{X^{2}+(\cos{\theta})^{2}}}\label{A1}\\
\cos{\rho}\!=\!-\frac{\sqrt{X^{2}+1}\cos{\theta}}{\sqrt{X^{2}+(\cos{\theta})^{2}}}\label{A2}
\end{eqnarray}
\begin{eqnarray}
\sinh{\alpha}\!=\!-\frac{\sin{\theta}}{\sqrt{X^{2}+(\cos{\theta})^{2}}}\label{R1}\\
\cosh{\alpha}\!=\!\frac{\sqrt{X^{2}+1}}{\sqrt{X^{2}+(\cos{\theta})^{2}}}\label{R2}
\end{eqnarray}
as well as
\begin{eqnarray}
\sin{\beta}\!=\!-\frac{\cos{\theta}}{\sqrt{X^{2}+(\cos{\theta})^{2}}}\\
\cos{\beta}\!=\!\frac{X}{\sqrt{X^{2}+(\cos{\theta})^{2}}}
\end{eqnarray}
in terms of $X\!=\!X(r,\theta)$ in general: case $X\!=\!\sqrt{1\!-q^{4}}/q^{2}$ gives the hydrogen atom \cite{Fabbri:2018crr} while $X\!=\!(a^{2}\!-\!r^{2})(2ar)^{-1}$ is the harmonic oscillator \cite{Fabbri:2019kyd}. However, in general $X$ is a not yet specified function, the only one that has to be found as solution of the field equations. The substitution of $\beta$, $\alpha$ and $\rho$ into (\ref{a}-\ref{d}) gives
\begin{eqnarray}
\nonumber
&r\partial_{r}\zeta\!+\!(\tan{\theta}\tanh{\zeta}\partial_{\theta}\zeta-1)+\\
&+2(\varepsilon+E)r\cosh{\zeta}+2L\!-\!2mr\sinh{\zeta}\!=\!0\label{A}\\
\nonumber
&r\partial_{r}\zeta\!-\!(\cot{\theta}\coth{\zeta}\partial_{\theta}\zeta+1)+\\
&+2(\varepsilon+E)r\cosh{\zeta}+2L\!-\!2mr\sinh{\zeta}\!=\!0\label{B}\\
\nonumber
&[(\sinh{\zeta})^{2}\!+\!(\cos{\theta})^{2}]r\partial_{r}\nu
\!+\!2mr(\cos{\theta})^{2}\cosh{\zeta}-\\
\nonumber
&-(\partial_{\theta}\zeta\cos{\theta}\sin{\theta}+\sinh{\zeta}\cosh{\zeta})+\\
&+[2(\varepsilon+E)r\sin{\theta}\!+\!2L\frac{\cosh{\zeta}}{\sin{\theta}}]
\sinh{\zeta}\sin{\theta}\!=\!0\label{C}\\
\nonumber
&[(\sinh{\zeta})^{2}\!+\!(\cos{\theta})^{2}]\partial_{\theta}\nu
\!-\!2mr\sinh{\zeta}\cos{\theta}\sin{\theta}+\\
\nonumber
&+\cos{\theta}\sin{\theta}r\partial_{r}\zeta+\\
&+[2(\varepsilon+E)r\sin{\theta}\!+\!2L\frac{\cosh{\zeta}}{\sin{\theta}}]
\cosh{\zeta}\cos{\theta}\!=\!0\label{D}
\end{eqnarray}
with $X\!=\!\sinh{\zeta}$ as well as $\ln{(\phi^{2}r^{2}\sin{\theta})}\!=\!\nu$ for simplicity.

Solutions will be taken as square-integrable, that is
\begin{eqnarray}
\!\!\!\!\!\!\!\!I\!=\!\!\int\!\!\!\!\int\!\!\!\!\int_{\Omega}\!\!\phi^{2}d\Omega
\!=\!2\pi\!\int_{0}^{\infty}\!\!\!\!\int_{0}^{\pi}\!\!\!\!\phi^{2}r^{2}\sin{\theta}d\theta dr
\label{I}
\end{eqnarray}
must be finite, or in other words convergent.

For the integration, it is easy to see that after combining (\ref{A}) and (\ref{B}) one gets 
\begin{eqnarray}
&\zeta\!=\!\zeta(r)\label{zeta}
\end{eqnarray}
such that
\begin{eqnarray}
&\zeta'\!+\!2(\varepsilon\!+\!E)\cosh{\zeta}\!+\!(2L\!-\!1)/r\!-\!2m\sinh{\zeta}\!=\!0
\end{eqnarray}
and plugging them into (\ref{C}) and (\ref{D}) we get
\begin{eqnarray}
\nu\!=\!\ln{\left[\frac{\sqrt{(\sinh{\zeta})^{2}
\!+\!(\cos{\theta})^{2}}}{(\sin{\theta})^{2L}}\right]}\!+\!V(r)\label{nu}
\end{eqnarray}
such that
\begin{eqnarray}
&V'\!=\!-2m\cosh{\zeta}\!+\!2(\varepsilon\!+\!E)\sinh{\zeta}
\end{eqnarray}
whose solution would give the module. The equations to solve are therefore
\begin{eqnarray}
&\!\!\!\!\!\!\!\!Z'
\!+\!(\varepsilon\!+\!E\!-\!m)Z^{2}\!+\!(2L\!-\!1)Z/r\!+\!(\varepsilon\!+\!E\!+\!m)\!=\!0
\label{ext}\\
&V'\!=\!(\varepsilon\!+\!E\!-\!m)Z\!-\!(\varepsilon\!+\!E\!+\!m)/Z\label{extV}
\end{eqnarray}
where we have set $\zeta\!=\!\ln{Z}$ for simplicity.

As is clear, the trial solution permitted the integration of the angular dependence. At this stage one can already see that to have the convergence of $\phi^{2}$ as defined in (\ref{spinorch}) a necessary condition is to have $L\!<\!1/2$ strictly. To see this just plug the expression for $\phi^{2}$ in the volume integral (\ref{I}) and see that because of (\ref{nu}) we have
\begin{eqnarray}
\!\!\!\!\!\!\!\!I\!=\!2\pi\!\int_{0}^{\infty}\!\!\!\!
\int_{0}^{\pi}\frac{\sqrt{(\sinh{\zeta})^{2}
\!+\!(\cos{\theta})^{2}}}{(\sin{\theta})^{2L}}e^{V}d\theta dr
\end{eqnarray}
where $\zeta$ and $V$ are both function of $r$ alone and they are still undetermined. However, because $(\cos{\theta})^{2}\!\leqslant\!1$ then
\begin{eqnarray}
\!\!\!\!\!\!\!\!I\!\leqslant\!2\pi\!\int_{0}^{\pi}\frac{d\theta }{(\sin{\theta})^{2L}}
\int_{0}^{\infty}\!\!\!\!\cosh{\zeta}\ e^{V}dr
\end{eqnarray}
where the integration in $r$ is still unknown but the integration in $\theta$ can be evaluated. In this case, to have any hope of convergence, the $\sin{\theta}$ function must be brought at the numerator, and this can only be possible for $L\!\leqslant\!0$ in general. Because, as we already said, $L$ is the angular momentum of the Dirac particle with $L\!=\!\pm\!1/2$ being its only possible values, it follows that $L\!=\!-1/2$ is the only value compatible with our assumptions. In addition, as we have stated above, $E\!=\!m$ will be chosen. As working hypothesis, we begin by studying the case $\varepsilon\!=\!-|\varepsilon|$ representing negative tensions, and thus attractive situations.

In this case, which we label with \emph{ex} for the reason that will become clearer later, (\ref{ext}) is a Riccati equation \cite{I} that after the Cole-Hopf transformation \cite{Y} like
\begin{equation}
Z_{\mathrm{ex}}\!=\!-\frac{z'_{\mathrm{ex}}}{z_{\mathrm{ex}}|\varepsilon|}
\end{equation}
results into
\begin{eqnarray}
&z_{\mathrm{ex}}''\!-\!2z_{\mathrm{ex}}'/r
\!-\!z_{\mathrm{ex}}|\varepsilon|(2m\!-\!|\varepsilon|)\!=\!0
\end{eqnarray}
so that fixing $|\varepsilon|<2m$ and so $|\varepsilon|(2m\!-\!|\varepsilon|)\!=\!H^{2}$ and then calling $rH\!=\!R$ we get
\begin{eqnarray}
&z_{\mathrm{ex}}\!=\!Ke^{-R}(R\!+\!1)
\end{eqnarray}
as the only convergent solution. Therefore
\begin{equation}
Z_{\mathrm{ex}}\!=\!R(R\!+\!1)^{-1}\sqrt{2m/|\varepsilon|\!-\!1}
\end{equation}
so that $Z_{\mathrm{ex}}\!>\!0$ as it is supposed to be and
\begin{equation}
X_{\mathrm{ex}}\!=\!\frac{2R^{2}(m\!-\!|\varepsilon|)\!-\!|\varepsilon|(2R\!+\!1)}{2HR(R\!+\!1)}
\end{equation}
for the unknown function. Thus
\begin{eqnarray}
&\beta_{\mathrm{ex}}\!=\!-\arctan{\left(\cos{\theta}/X_{\mathrm{ex}}\right)}\\
&\phi^{2}_{\mathrm{ex}}\!=\!\Gamma^{2}\left(1+R\right)R^{-3}e^{-2R}\sqrt{X_{\mathrm{ex}}^{2}
\!+\!(\cos{\theta})^{2}}
\end{eqnarray}
where $\Gamma$ is a generic integration constant. Clearly, a large value of $R$ gives a behaviour for which the volume integral of $\phi^{2}_{\mathrm{ex}}$ converges toward infinity. However, there clearly is a divergence for $R\!\rightarrow\!0$ rendering the volume integral of $\phi^{2}_{\mathrm{ex}}$ divergent in the origin. This means that this solution should be seen as an external solution. And this is why this solution has been indicated with the label \emph{ex} above.

This choice however is not the only possibility, and it is possible to consider the alternative trial solution which is obtained by considering (\ref{A1}-\ref{R2}) unchanged but with
\begin{eqnarray}
\sin{\beta}\!=\!\frac{\cos{\theta}}{\sqrt{X^{2}+(\cos{\theta})^{2}}}\\
\cos{\beta}\!=\!-\frac{X}{\sqrt{X^{2}+(\cos{\theta})^{2}}}
\end{eqnarray}
as a new type of Yvon-Takabayashi angle. Now the Dirac equation gives the same (\ref{zeta}, \ref{nu}) as above but also
\begin{eqnarray}
&\!\!\!\!\!\!\!\!Z'
\!+\!(\varepsilon\!+\!E\!+\!m)Z^{2}\!+\!(2L\!-\!1)Z/r\!+\!(\varepsilon\!+\!E\!-\!m)\!=\!0
\label{int}\\
&V'\!=\!(\varepsilon\!+\!E\!+\!m)Z\!-\!(\varepsilon\!+\!E\!-\!m)/Z\label{intV}
\end{eqnarray}
where we have defined $Z$ again as we have done above.

Such a trial solution also permits the integration of the angular dependence. And again as above we will specify to the $L\!=\!-1/2$ case. And similarly, $E\!=\!m$ will also be chosen. For this second solution however, we set $\varepsilon\!=\!0$ as constraint specifying the total lack of space-time tension.

In this case, labelled \emph{in} because, complementary to the above, this is the internal solution, (\ref{int}) is a Riccati equation that after the new Cole-Hopf transformation that is now given by
\begin{equation}
Z_{\mathrm{in}}\!=\!\frac{z_{\mathrm{in}}'}{2mz_{\mathrm{in}}}
\end{equation}
it results into
\begin{eqnarray}
&z''_{\mathrm{in}}\!-\!2z_{\mathrm{in}}'/r\!=\!0
\end{eqnarray}
which admits the solution
\begin{eqnarray}
&z_{\mathrm{in}}\!=\!Kr^3
\end{eqnarray}
as the only solution regular at the origin. Then
\begin{equation}
Z_{\mathrm{in}}\!=\!3/\widetilde{R}
\end{equation}

\

\noindent having written $2mr\!=\!\widetilde{R}$ for simplicity and
\begin{equation}
X_{\mathrm{in}}\!=\!\frac{9\!-\!\widetilde{R}^{2}}{6\widetilde{R}}
\end{equation}
for the unknown. Therefore
\begin{eqnarray}
&\beta_{\mathrm{in}}\!=\!-\arctan{\left(\cos{\theta}/X_{\mathrm{in}}\right)}\\
&\phi^{2}_{\mathrm{in}}\!=\!Q^{2}\widetilde{R}\sqrt{X_{\mathrm{in}}^{2}\!+\!(\cos{\theta})^{2}}
\end{eqnarray}
where $Q$ is a generic integration constant. Now, for small values of $\widetilde{R}$ the volume integral of $\phi^{2}_{\mathrm{in}}$ converges close to the origin. Conversely, the volume integral of $\phi^{2}_{\mathrm{in}}$ fails to converge at infinity. This means that this solution must be seen as an internal, as indicated. These two solutions have two complementary behaviours whether they are at infinity or close to the origin of the coordinate system.

The explicit expression of the physical observables is
\begin{eqnarray}
\beta_{\mathrm{in}}\!=\!-\arctan{\left(\frac{12mr\cos{\theta}}{9\!-\!4m^{2}r^{2}}\right)}\\
\phi^{2}_{\mathrm{in}}\!=\!\frac{1}{6}Q^{2}\sqrt{(9\!-\!4m^{2}r^{2})^{2}\!+\!(12mr\cos{\theta})^{2}}
\end{eqnarray}
for the internal solution, and that is the solution defined in any ball centred in the origin, and
\begin{widetext}
\begin{eqnarray}
\beta_{\mathrm{ex}}\!=\!-\arctan{\left[\frac{2r(2m\!-\!|\varepsilon|)
\left(r\sqrt{|\varepsilon|(2m\!-\!|\varepsilon|)}\!+\!1\right)
\cos{\theta}}
{2r^{2}(2m\!-\!|\varepsilon|)(m\!-\!|\varepsilon|)
\!-\!2r\sqrt{|\varepsilon|(2m\!-\!|\varepsilon|)}\!-\!1}\right]}\\
\nonumber
\phi^{2}_{\mathrm{ex}}\!=\!\Gamma^{2}e^{-2r\sqrt{|\varepsilon|(2m-|\varepsilon|)}}
\left[\frac{1+r\sqrt{|\varepsilon|(2m\!-\!|\varepsilon|)}}
{2r^{4}\sqrt{|\varepsilon|^{3}(2m\!-\!|\varepsilon|)^{5}}
\left(r\sqrt{|\varepsilon|(2m\!-\!|\varepsilon|)}\!+\!1\right)}\right]\cdot\\
\cdot\sqrt{[2r^{2}(2m\!-\!|\varepsilon|)(m\!-\!|\varepsilon|)
\!-\!2r\sqrt{|\varepsilon|(2m\!-\!|\varepsilon|)}\!-\!1]^{2}
\!+\!\left[2r(2m\!-\!|\varepsilon|)\left(r\sqrt{|\varepsilon|(2m\!-\!|\varepsilon|)}\!+\!1\right)
\cos{\theta}\right]^{2}}
\end{eqnarray} 
\end{widetext}
for the external solution, and that is the solution defined in all the space external to any ball centred in the origin.

These two exact solutions are related to the two possible different values of the tension $|\varepsilon|$ according to whether it is zero or not. The non-zero value, giving the behaviour at infinity, can be justified by generality. The null value, giving the behaviour at the origin, can instead be justified with more difficulty since we might always wonder why of all possible values the very peculiar $|\varepsilon|\!=\!0$ should in fact be selected. The answer lies in physical considerations.

If we allowed only the exterior solution, near the origin it would make all volume integrals diverge, including the spin and energy density of the field. Thus near the origin we would no longer be able to maintain the approximation of vanishing torsion and curvature, and therefore the problem would become that of finding the solution of an interacting theory. Alas, this problem is very difficult to solve, but we do not always need to find exact solutions in order to know properties of the matter distribution. In fact, as it has been discussed in \cite{Fabbri:2017rjf}, a spinor field in its own torsion-gravity behaves in such a way that at small scales the spin-torsion interaction provides the dominant negative potential that reverts the sign of the curvature around a particle hence averting gravitational singularity formation. In such case, the space-time would behave as if it had no effect, and the value $|\varepsilon|\!=\!0$ becomes clear.
%%%%%%%%%%%%%%%%%%%%%%%%%%%%%%%%%%%%%%%%%%%%%%%%%%%%%%%%%%%%%%%%%%%%%%%%%%%%%%%%%%%%%%%%%%%%%%%%%%%
%%%%%%%%%%%%%%%%%%%%%%%%%%%%%%%%%%%%%%%%%%%%%%%%%%%%%%%%%%%%%%%%%%%%%%%%%%%%%%%%%%%%%%%%%%%%%%%%%%%
\section{Junction Conditions}
While the two solutions found above correspond to two different values of the tension of the tensorial connection, it is nevertheless possible to have them combined into the single solution obtained whenever $\phi_{\mathrm{in}}\!=\!\phi_{\mathrm{ex}}$ and $\beta_{\mathrm{in}}\!=\!\beta_{\mathrm{ex}}$ in a given boundary $r\!=\!b$ then providing the conditions
\begin{eqnarray}
\frac{2b^{2}(2m\!-\!|\varepsilon|)(m\!-\!|\varepsilon|)\!-\!2Hb\!-\!1}
{(2m\!-\!|\varepsilon|)(Hb\!+\!1)}\!=\!\frac{9\!-\!4m^{2}b^{2}}{6m}\label{discr}\\
2H^{3}b^{4}me^{2Hb}(1\!+\!Hb)^{-1}\!=\!\Gamma^{2}/Q^{2}\label{con}
\end{eqnarray}
where $b$ is the radius of the boundary where internal and external solutions are defined. Clearly, the second condition fixes the ratio of the two integration constants, while the first condition is a constraint between the parameters.

It is important to notice that (\ref{discr}) is just the expression of the condition $X_{\mathrm{in}}\!=\!X_{\mathrm{ex}}$ and it implies that $\alpha$ and $\rho$ and therefore frames and co-frames are also continuous.

We remark that since the Dirac equation is relativistic, and thus of the first order derivative, it requires continuity of the solution, but not of its derivatives. However, a look at equations (\ref{ext}, \ref{extV}) or (\ref{int}, \ref{intV}) makes it clear that continuity of $X$ implies also the continuity of $X'$ and $V'$ and therefore the continuity of the derivatives of the solutions. This boot-strap process is precisely due to the first order in the derivatives of the Dirac equations.

Compared to the non-relativistic case giving discretization on the energy, the relativistic case gives discretization of the mass. That is to say, once we give the external conditions on the region of radius $b$ where the tension $\varepsilon$ of the tensorial connection is not zero, relationship (\ref{discr}) provides a constraint resulting in the discretization of the mass spectrum. In a general case, writing (\ref{discr}) in terms of $k\!=\!b|\varepsilon|$ and $x\!=\!m/|\varepsilon|$ and then $\sqrt{2x\!-\!1}\!=\!y$ gives 
\begin{eqnarray}
&(ky^{3}\!+\!4y^{2}\!+\!ky\!+\!1)(k^{2}y^{4}\!+\!k^{2}y^{2}\!-\!3ky\!-\!3)\!=\!0
\end{eqnarray}
and because $k$ and $y$ are always positive the first factor has no positive solution. As for the second factor, it can always be solved by employing the quartic formula, although the explicit solutions are much too complicated to be insightful.\! But plotting the solution for specific values of $k$ it is easy to see that in general two solutions are always complex and one of the real ones is always negative, so that we have a single real and positive solution. Large values of $k$ give a $y$ that goes to zero and hence the mass tends to its limiting value $|\varepsilon|/2$ whereas small values of $k$ give a $y$ larger and larger and thus a mass that is larger and larger. The special case $k\!=\!1$ gives $y\!\approx\!1.514$ and so $m\!\approx\!1.646 |\varepsilon|$ after a quick numerical evaluation. It is interesting to see that in this case the condition $b|\varepsilon|\!=\!1$ can be interpreted by saying that the radius $b$ is of the order of the Compton length of the tension $|\varepsilon|$ of the tensorial connection, which is what one might have expected.
%%%%%%%%%%%%%%%%%%%%%%%%%%%%%%%%%%%%%%%%%%%%%%%%%%%%%%%%%%%%%%%%%%%%%%%%%%%%%%%%%%%%%%%%%%%%%%%%%%%
%%%%%%%%%%%%%%%%%%%%%%%%%%%%%%%%%%%%%%%%%%%%%%%%%%%%%%%%%%%%%%%%%%%%%%%%%%%%%%%%%%%%%%%%%%%%%%%%%%%
\section{Finite Integral}
To check consistency, we assess the square-integrability by computing the volume integral (\ref{I}) as
\begin{eqnarray}
\!\!\!\!I\!=\!2\pi\!\!\int_{0}^{b}\!\!\int_{0}^{\pi}\!\!\!\!\phi^{2}_{\mathrm{in}}r^{2}\sin{\theta}d\theta dr
\!+\!2\pi\!\!\int_{b}^{\infty}\!\!\!\!\int_{0}^{\pi}\!\!\!\!\phi^{2}_{\mathrm{ex}}r^{2}\sin{\theta}d\theta dr
\end{eqnarray}
to be evaluated in the internal and external regions. The exact evaluation is rather long, but because $|\!\cos{\theta}|\!\leqslant\!1$ it is possible to overestimate the integral with one that has variable separation and therefore it is easier. Evaluations are again long in general, but it is possible to see that for the value $k\!=\!1$ we have that $I\!<\!17\pi Q^{2}$ which is finite.

As usual, it would be possible to employ the finite value of the volume integral to normalize the solution, thus fixing the only constant that remained free in the problem.

In \cite{Fabbri:2020ypd} our motivation was to find solutions localized at infinity and regular in the origin, with junction conditions giving discretization of the mass spectrum. In the present paper the search for solutions of this type was empowered by the fact that the solutions we found are more general than that of \cite{Fabbri:2020ypd} as (\ref{A1}-\ref{R2}) are less constraining than the conditions assumed in the aforementioned paper.
%%%%%%%%%%%%%%%%%%%%%%%%%%%%%%%%%%%%%%%%%%%%%%%%%%%%%%%%%%%%%%%%%%%%%%%%%%%%%%%%%%%%%%%%%%%%%%%%%%%
%%%%%%%%%%%%%%%%%%%%%%%%%%%%%%%%%%%%%%%%%%%%%%%%%%%%%%%%%%%%%%%%%%%%%%%%%%%%%%%%%%%%%%%%%%%%%%%%%%%
\section{Conclusion}
In this paper we found two exact solutions corresponding to two different cases of $|\varepsilon|$ being it zero or not. The two solutions have opposite properties of convergence for their volume integrals whether they are close to the origin or at infinity, and so we have joined the two well-behaved branches into a single solution at a given radius $b$ asking for continuity. Such a continuity gives rise to conditions on the mass that result into a single possible value given in terms of $|\varepsilon|b$ and which is reminiscent of the conditions that fix the values of the energy levels in non-relativistic version of quantum mechanics and quantum field theory.

As clear, more general solutions should be found, and our method for the integration of the angular and radial variables expressed by assuming conditions (\ref{A1}-\ref{R2}) may be helpful in finding solutions in more general situations.

It is also important to remark that our trial solution is best seen when working with spinors in polar form.

\

The data that support the findings of this study are available from the corresponding author upon request.
%%%%%%%%%%%%%%%%%%%%%%%%%%%%%%%%%%%%%%%%%%%%%%%%%%%%%%%%%%%%%%%%%%%%%%%%%%%%%%%%%%%%%%%%%%%%%%%%%%%
%%%%%%%%%%%%%%%%%%%%%%%%%%%%%%%%%%%%%%%%%%%%%%%%%%%%%%%%%%%%%%%%%%%%%%%%%%%%%%%%%%%%%%%%%%%%%%%%%%%

%%%%%%%%%%%%%%%%%%%%%%%%%%%%%%%%%%%%%%%%%%%%%%%%%%%%%%%%%%%%%%%%%%%%%%%%%%%%%%%%%%%%%%%%%%%%%%%%%%%
\end{document}